\def\cm{{\rm\thinspace cm}}

\def\eV{{\rm\thinspace eV}}

\def\keV{{\rm\thinspace keV}}

\def\Msun{\hbox{$\rm\thinspace M_{\odot}$}}

\def\pcmsq{\hbox{$\cm^{-2}\,$}}

\def\psqcm{\hbox{$\cm^{-2}\,$}}

\documentclass[12pt,a4paper]{article}
\usepackage{psfig}
\usepackage{times}
\usepackage{citesupernumber}
\usepackage[margin=1in]{geometry}
\usepackage{setspace}

\title{The detection of Broad Iron K and L line emission in the
  Narrow-Line Seyfert 1 Galaxy 1H\,0707-495 using XMM-Newton}

\begin{document}

\date{}

\maketitle 

\author{\noindent A.C. Fabian$^1$, A. Zoghbi$^1$, R.R. Ross$^2$,
  P. Uttley$^3$, L.C. Gallo$^4$, W.N. Brandt$^5$, A. Blustin$^1$,
  T. Boller$^6$, M.D. Caballero-Garcia$^1$, J. Larsson$^1$,
  J.M. Miller$^7$, G. Miniutti$^8$, G. Ponti$^9$, R.C. Reis$^1$,
  C.S. Reynolds$^{10}$, Y. Tanaka$^6$ \& A.J. Young$^{11}$\\\\
  \small{1. Institute of Astronomy, Madingley Road, Cambridge CB3 0HA, UK\\
    2. Physics Department, College of the Holy Cross, Worcester, MA 01610, USA\\
    3. School of Physics and Astronomy, University of Southampton,
    Highfield,    Southampton SO17 1BJ, UK\\
    4. Department of Astronomy and Physics, Saint Mary's University,
    Halifax, Nova Scotia, Canada\\
    5. Department of Astronomy and Astrophysics, The Pennsylvania
    State University, 525 Davey Lab, University Park, PA 16802 USA\\
    6. Max-Planck-Institut für Extraterrestrische Physik,
    Giessenbachstraße, Postfach 1312, 85741 Garching, Germany \\
    7. Department of Astronomy, University of Michigan, Ann Arbor MI
    48109, USA\\
    8. Laboratorio de Astrof\'isica Espacial y F\'isica Fundamental
    (CSIC--INTA), P.O. Box 78, E--28691, Villanueva de la Ca\~{n}ada,
    Madrid, Spain\\
    9. Laboratoire APC, UMR 7164, 10 rue A. Domon et L. Duquet, 75205
    Paris, France\\
    10. Department of Astronomy and the Center for Theory and
    Computation, University of Maryland, College
    park, MD20742, USA\\
    11. H. H. Wills Physics Laboratory, University of Bristol, Tyndall
    Avenue, Bristol
    BS8 1TL\\
  }

\begin{spacing}{2} 
   
  {\bf Since the discovery of the first broad iron-K line in 1995 from
    the Seyfert Galaxy MCG--6-30-15$^{1}$, broad iron-K lines have
    been found in several other Seyfert galaxies$^{2 }$, from
    accreting stellar mass black holes$^{3}$ and even from accreting
    neutron stars$^{4}$. The iron-K line is prominent in the
    reflection spectrum$^{5,6}$ created by the hard X-ray continuum
    irradiating dense accreting matter. Relativistic distortion$^7$ of
    the line makes it sensitive to the strong gravity and spin of the
    black hole $^8$. The accompanying iron-L line emission should be
    detectable when the iron abundance is high.  Here we report the
    first discovery of both iron-K and L emission, using XMM-Newton
    observations of the Narrow-Line Seyfert 1 Galaxy$^9$ 1H\,0707-495.
    The bright Fe-L emission has enabled us, for the first time, to
    detect a reverberation lag of 30 s between the direct X-ray
    continuum and its reflection from matter falling into the hole.
    The observed reverberation timescale is comparable to the
    light-crossing time of the innermost radii around a supermassive
    black hole.  The combination of spectral and timing data on
    1H\,0707-495 provides strong evidence that we are witnessing
    emission from matter within a gravitational radius, or a fraction
    of a light-minute, from the event horizon of a rapidly-spinning,
    massive black hole.}

  The galaxy 1H0707-495 has been observed several times by
  XMM-Newton$^{10-12}$. The first observation revealed a sharp and
  deep spectral drop at 7~keV in the rest frame (the source redshift
  is 0.041) but no narrow emission features. This led to two main
  interpretations$^{10}$: either the source is partially obscured by a
  large column of iron-rich material or it has very strong X-ray
  reflection$^{11}$ in its innermost regions where relativistic
  effects modify the observed spectrum. In other words the sharp drop
  is either due to a photoelectric absorption edge, or the blue wing
  of a line partially shaped by relativistic Doppler shifts. The
  absorption origin requires that the iron abundance is about 30 times
  the Solar value (unless the spectrum is e-folding below 10
  keV$^{13}$, when the value is reduced), whereas reflection requires
  that this factor is between 5 and 10. The extreme variability of the
  source moreover appears to be due to changes mostly in the intensity
  of the powerlaw continuum, above 1 keV, which does not make sense in
  a partial covering model.

  We have analysed the spectral variability of new observations of the
  source taken with XMM-Newton in January 2008. Variations from 1 to
  12~ct~s$^{-1}$ are seen over 4 XMM orbits. The difference spectrum
  between low and high flux states is well fitted by a powerlaw
  continuum with photon index $\Gamma=3$, with an excess at lower
  (below 1.1~keV) energies. Strong skewed and broad residuals are seen
  above such a powerlaw fitted to the full source spectrum, peaking
  around 0.9 and 6.7 keV. The data are well described by a simple
  phenomenological model composed of a powerlaw continuum, a soft
  blackbody, two relativistically-broad (Laor$^{14}$) lines and
  Galactic absorption (corresponding to $N_{\rm H}=5\times
  10^{20}\psqcm$). We show the ratio of the spectrum to the continuum
  in that model in Fig.~1. This ratio spectrum is clearly dominated by
  two strong broad emission lines. They are characterised by energies
  of 0.89 and 6.41~keV (in our frame), innermost radius of $1.3 r_{\rm
    g}$ ($1r_{\rm g}=GM/c^2$), outermost radius of $400r_{\rm g}$, an
  emissivity index of 4 and an inclination of 55.7 deg. The
  normalizations of the lines (in photon spectra) are in the ratio of
  20:1 and their rest energies correspond well to ionized iron-L and
  K, respectively. We have tried a variety of continuum models and
  always obtain similar results.  The data have also been fitted with
  a self-consistent model reflection spectrum$^{15}$ (Fig.~2; see
  Supplementary Information for more details) which reproduces the
  correct relative fluxes for the lines. Iron is almost 9 times the
  Solar abundance value with the other elements at the Solar
  values. Perhaps a dense nuclear star cluster has led to the
  formation of massive white dwarf binaries which have enriched the
  nucleus with SN Ia ejecta rich in iron (such a scenario has been
  invoked in globular star clusters$^{16}$).

  The presence of broadened and skewed lines other than iron-K is an
  important prediction of the disk reflection model that we have now
  confirmed.  Furthermore, we have shown that the relative strengths
  of the observed iron-L and iron-K lines agree well with predictions
  based on atomic physics.  While it is possible to construct a
  partial-covering, absorption-dominated, model$^{10,12}$ for the
  prominent K-shell iron feature, the L-shell absorption edge of
  ionized iron required around 1~keV is accompanied by an unacceptably
  strong absorption feature around 0.75~keV from an unresolved
  transition array of Fe IX-XI. Previous work$^{10,12}$ on absorption
  models for 1H\,0707-495 have been unable to account for the spectral
  structure around 1~keV without invoking emission. The strong
  variability, and shape, of this emission component points to the
  inner regions of the flow and thus to a reflection solution. 
 
  The source fractional rms variability is roughly constant below
  1~keV, increases abruptly at 1~keV and then drops back to the soft
  level above 4~keV, in agreement with the expectation of the
  two-component model (reflection plus powerlaw) used to fit the
  data. Both components vary in amplitude but the powerlaw nearly
  twice as much. The non-linear behaviour of the variability in
  accreting black holes$^{17}$ is considered to be due to the
  cumulative random effects of the orbital variations from many radii
  effectively multiplied together. On the shortest time-scales
  light-crossing effects will become important, since the light path
  for the direct primary radiation is shorter than that for the
  reflected radiation. This effect is seen for the first time in the
  frequency-dependent lags, Fig.~3. The large positive lag for
  variations slower than 0.6 mHz (time-scales greater than 30 min) is
  probably due to the inward drift of accretion fluctuations through
  the emitting region$^{18}$ , causing the density and thus ionization
  state of the irradiated disc to respond first. Variations faster
  than 0.6mHz show a negative lag in the sense that the soft
  reflection dominated band follows the hard powerlaw dominated one by
  about 30~s. This is in the opposite sense to a Comptonization lag
  produced by upscattering of photons (or to any model where the
  spectral drop at 1~keV is instead produced by absorption which is
  responding to the changing continuum) and is explained by
  reverberation. If the lag time corresponds to the natural length of
  about $hr_{\rm g}$, we expect $h$ to be 2--5, then we deduce
  a mass for the black hole of about $7\times 10^6 h^{-1}\Msun$, which
  is reasonable for this source (no definitive mass is known; see e.g.
  ref 19) and implies that the accretion is (just) sub-Eddington. The
  breadth of the iron lines implies (using the methods of ref.~8) that
  the black hole has a high spin, of dimensionless spin parameter
  $a=cJ/M^2>0.98,$ and so much of the emission should originate from
  within a few gravitational radii.  Iron L-line emission should be
  detectable in similar sources$^{20}$ with high iron abundances,
  thereby enabling reverberation studies to me made.

\section{Acknowledgements}

ACF thanks the Royal Society for support. This work is based on
observations made with {\it XMM-Newton}, an ESA science mission with
instruments and contributions directly funded by ESA member states and
the USA (NASA). AZ acknowledges the support of the Algerian Higher
Education Ministry and STFC. CSR, WNB GM, GP and both RCR and AJB
acknowledge support from the US National Science Foundation, NASA, the
Spanish Ministerio de Ciencia e Innovaci\'on, Italian ANR and UK STFC
for support, respectively.

\end{spacing} 

\begin{figure}[h]
\centerline{\psfig{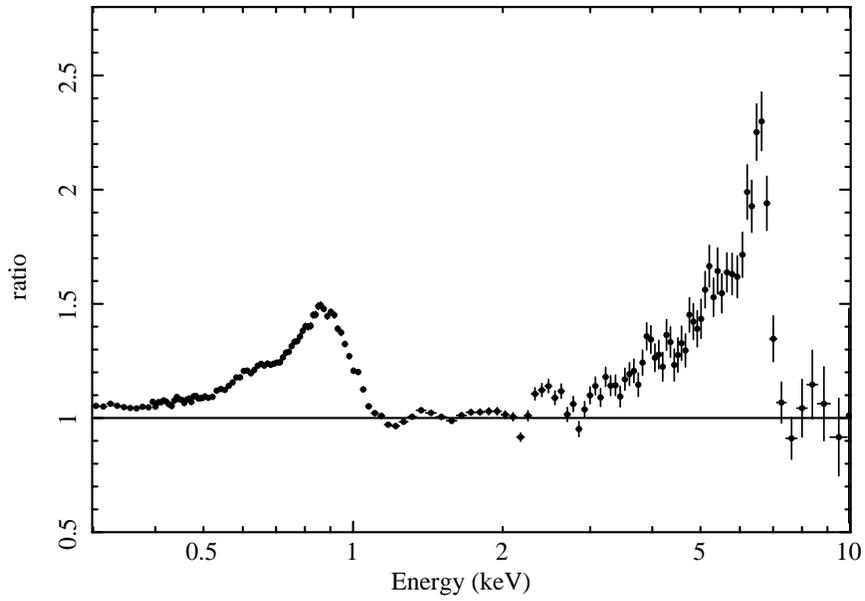}}
\caption{Ratio of the spectrum, obtained by combining all 4 XMM orbits,
  to a simple phenomenological model composed of a power-law, blackbody
  and 2 broad emission lines. The normalizations of the broad lines
  have been set to zero to make this plot. Ionized iron-L and K peak
  in the rest frame around 0.9~keV and 6.5-6.7~keV with equivalent
  widths of 180 and 970~eV, respectively.}
\end{figure}

\begin{figure}[h]
\centerline{\psfig{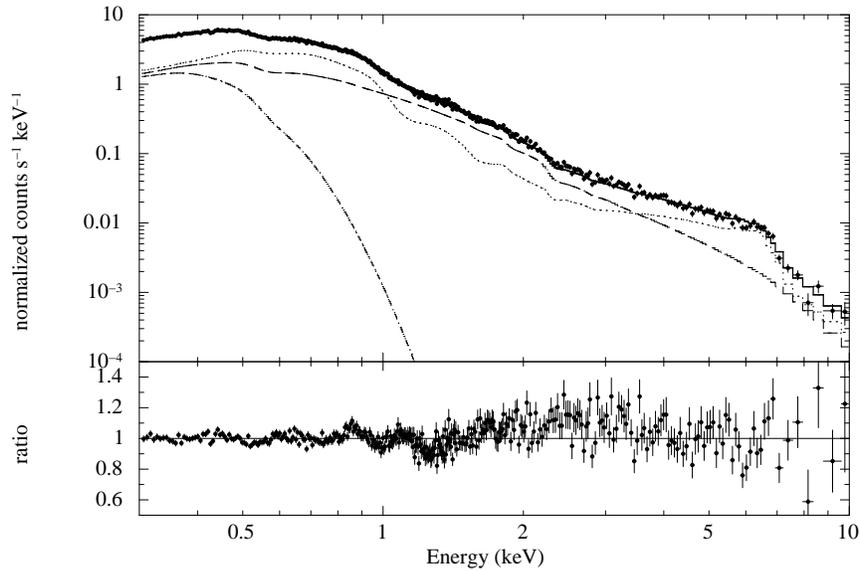}}
\caption{Spectrum of the first orbit showing the best-fitting,
  self-consistent, relativistically-blurred, reflection model. The
  total exposure time for this spectrum is 102~ks with the source
  changing flux continuously and by more than a factor of 2 every few
  ks. The offset in the 1.5--4~keV band is due to the simplicity of
  using just two components in the 1--10~keV band despite the high
  variability of the source. The addition of a further reflection
  component with higher ionization parameter considerably reduces this
  offset (see the Supplementary Information for more details). This is
  to be expected if the ionization changes with time or flux. The
  contributions of the power-law, reflection and blackbody components
  are indicated by the dashed, dotted and dash-dotted lines
  respectively.}
\end{figure}

\begin{figure}[h]
\centerline{\psfig{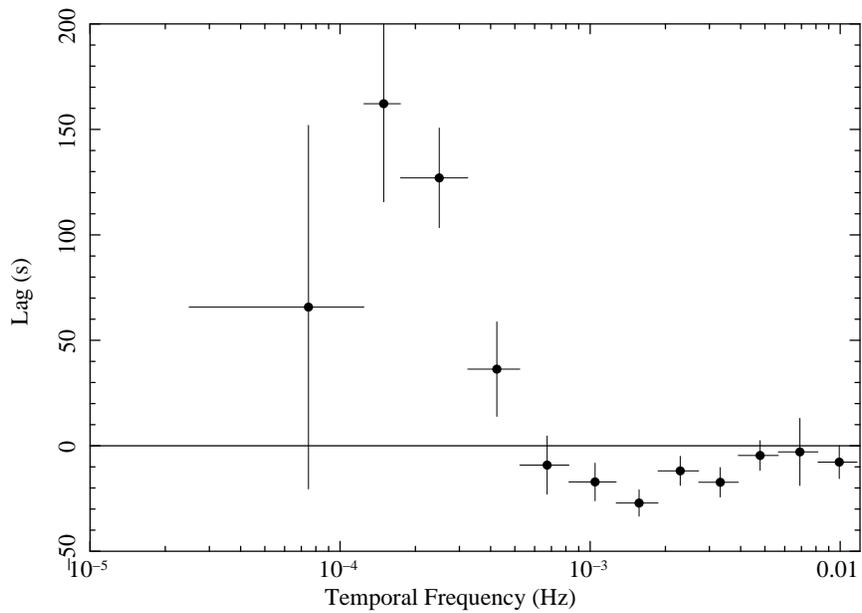}}
\caption{Frequency-dependent lags between the 0.3--1 and 1--4 keV
  bands. A negative lag, such as found above frequencies of $7\times
  10^{-4}$~Hz or timescales shorter than 30~min, indicates the harder
  flux, dominated by the powerlaw continuum, changing before the
  softer flux, which is dominated by reflection (particularly the iron
  L line). }
\end{figure}

\clearpage
\eject
\section{Supplementary Information}
1H0707-495 was observed for four consecutive orbits using XMM-Newton
starting on 2008, January 29th. The observations were made in the
large window imaging mode.  The data were analysed using XMM Science
Analysis System (SAS v8.0.0).  The light curve was filtered to remove
any background flares, and resulted in a total net exposure of 330 ks
for the EPIC PN camera.  No significant pileup was found in the
patterns, which were selected using PATTERN $\le 4$. Source spectra
were extracted from circular regions with a radius of 35 arcsec around
the source, and background spectra from regions on the same chip. The
spectra were then grouped to bins with a minimum of 20 counts
each. The response matrices were generated using RMFGEN and ARFGEN in
SAS. For the combined spectrum of the four orbits, the events files
were merged together before extracting the spetrum.

The light curve is shown (Fig.~4), together with the FeL line profile
as seen by the RGS (Fig.~6). No significant sharp absorption features
are seen around 1~keV in the RGS spectrum; Chandra HEG spectra have
tentatively shown features in this band (Leighly et al 2002). The
ratio plot for each orbit (Fig.~5), produced in a similar way to
Fig.~1, is also shown, as well as the difference spectrum
(Fig.~7). This spectrum fits a power-law above 1.1~keV and has a
significant excess at lower energies and a marginal excess from
4--7~keV; it is consistent within the error bars with the difference
spectrum in Ref.~11. We assume that the excesses are due to small
changes in the ionization parameter of the irradiated disc (sample
fits indicate that $\xi$ changes by about 20 per cent between
orbits). This also accounts for changes in the iron-L to K line ratio
between the orbits.

Fitting the data with a simple model consisting of 2 broad lines, a
power-law continnum and a low energy blackbody spectrum gives the
following parameters for the lines (assuming a Laor model and with
uncertainties at the 90\% confidence level): K-line energy (our frame)
$6.19^{+0.11}_{-0.14}\keV$, L-line energy
$0.864^{+0.010}_{-0.012}\keV,$ $r_{\rm in}=1.39^{+0.07}_{-0.01} r_{\rm
  g},$ disc inclination $55.7^{+0.6}_{-1.1}$ deg, emissivity index
$5.8^{+0.16}_{-0.26}$, $r_{\rm out}$ fixed at $400 r_{\rm g}$ and
photon index $ \Gamma=2.87^{+0.02}_{-0.01}$.

The best-fitting reflection model spectrum is shown (Fig.~8). This
spectrum has soft X-ray absorption due to a column of $8.7\times
10^{20}\pcmsq$ (the Galactic column density in that direction is
$5\times 10^{20}\pcmsq$ so we assume that there is some excess in the
host galaxy). The emission spectrum has a power-law (photon index of
3.09) plus low energy blackbody continuum ($kT=52\eV$) components and
a relativistically blurred reflection component (reflionx$^{15}$) with
iron at 8.88 times the Solar value and an ionization parameter of
53.4. The power-law has the same photon index as for the reflection
model, where it extends down to 0.1~keV. The blurring is for an inner
disc radius of $1.235r_{\rm g}$, and an emissivity index of 7.43
breaking at $4.32r_{\rm g}$ to an index of 1.93. The disc inclination
is 53.4 deg.  Note that the
blackbody component, presumably thermal disc emission, only makes a
significant contribution below 0.5~keV and has a temperature about one
third of that typically invoked for phenomenological models for AGN
(Crummy et al 2007 and references therein).

Adding a second reflection component with the same set of parameters
as the first, except for a higher ionization parameter in the range of
500--1200, gives a better fit with the residuals in the 1.5--4~keV
band (Fig.~10). Future work will address whether this is due to source
variability, to an intrinsic range of ionization, or to spatial
inhomogeneity of the reflector.  

Phenomenological absorption models, such as those discussed and fitted
above 2~keV to earlier data in Refs. 10 and 12, can fit the new
data. However the strong feature around 1~keV requires a broad
emission line, which implies reflection. An iron-L edge predicts other
features (UTAs) which are inconsistent with the data.

We plot the ratio of the data to a power-law continuum of photon
index 3 in Fig.~9 and the rms variability spectrum in Fig.~11. 

The mean high-frequency lag (above $7\times 10^{-4}$~Hz) between the
0.5--1 and 3--7~keV bands is $-6\pm18$s, indicating that the iron-L
and K features vary together, as expected for reflection.

The NLS1 IRAS\,13224-3809 shows a similar pair of FeK and FeL lines
(Ponti et al in prep).

\begin{figure}[h]
\centerline{\psfig{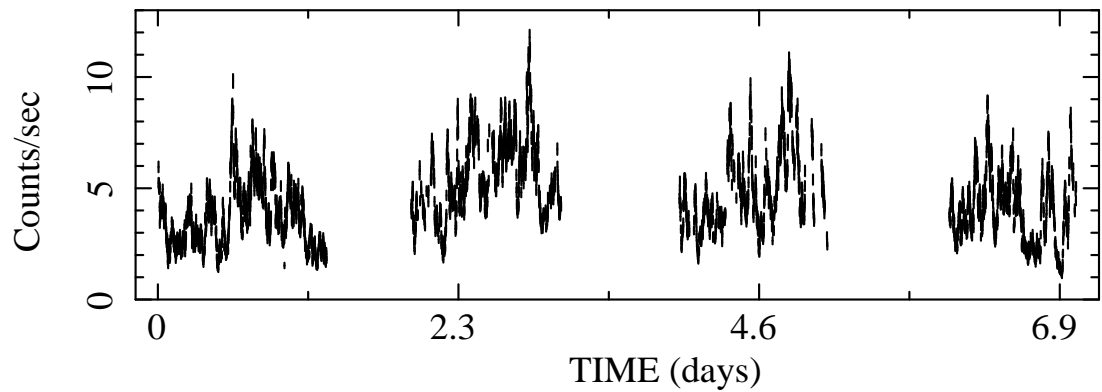}}
\caption{Total light curve (0.3--10~keV) of the XMM observation
  showing all 4 orbits.}
\end{figure}

\begin{figure}[h]
\centerline{\psfig{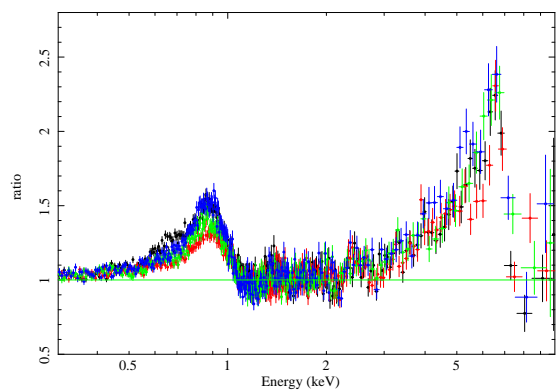}}
\caption{Ratio of the spectral data from each XMM orbit to a simple
  phenomological model composed of a power-law, blackbody and 2 broad
  emission lines. The normalizations of the broad lines have been set
  to zero to make this plot. Ionized iron-L and K peak  in the rest
  frame around 0.9~keV and  6.5-6.7~keV, respectively.}
\end{figure}

\begin{figure}[h]
\centerline{\psfig{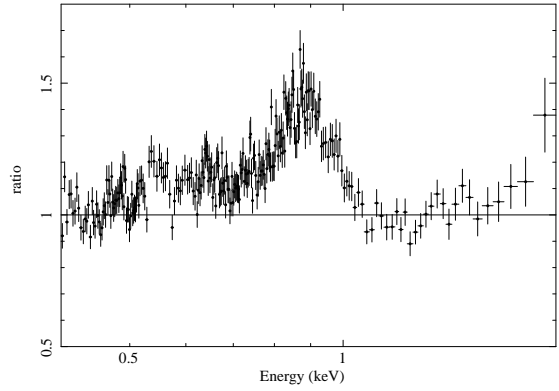}}
\caption{Ratio spectrum from the XMM reflection grating spectrometer
  (produced as for Fig. 1). We find no significant sharp features in
  this spectrum around 1~keV.}
\end{figure}

\begin{figure}[h]
\centerline{\psfig{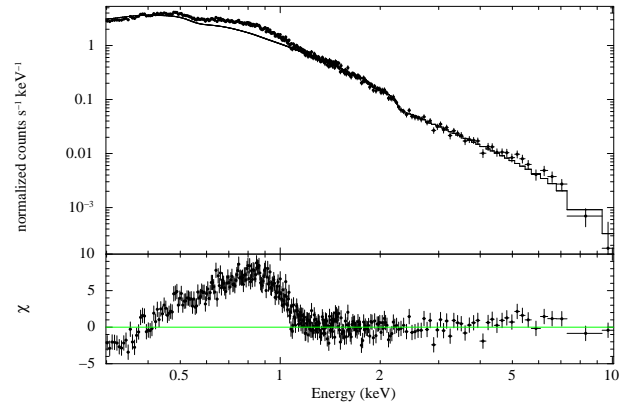}}
\caption{The difference spectrum for all 4 orbits (i.e. the spectrum
  formed by subtracting the fainter half of the spectra from the
  brighter half) fitted with a simple power-law spectrum with Galactic
  absorption. }
\end{figure}

\begin{figure}[h]
\centerline{\psfig{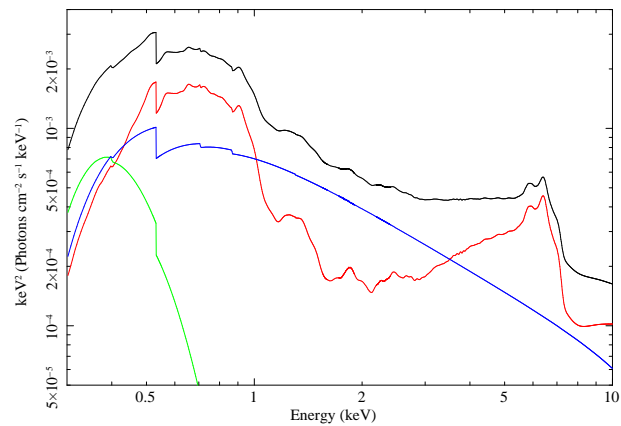}}
\caption{Best-fitting reflection model spectrum (see   text for
  details), with components shown.  }
\end{figure}

\begin{figure}[h]
\centerline{\psfig{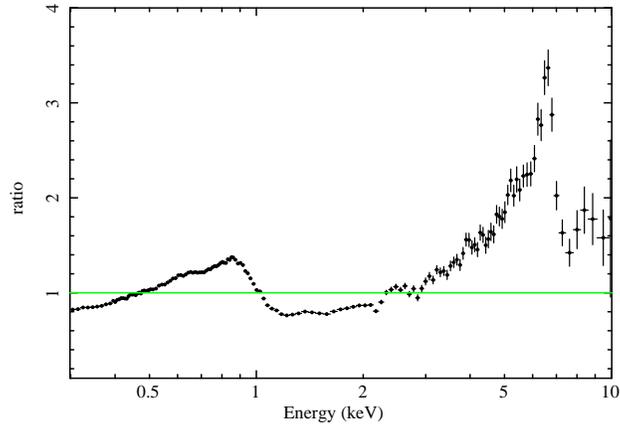}}
%\centerline{\psfig{width=0.5\textwidth,angle=-90,file=reflionx_notcheda.ps}}
\caption{ Ratio of the total spectrum to a powerlaw of
  photon index $\Gamma=3$. }
\end{figure}

\begin{figure}[h]
%\centerline{\psfig{width=0.5\textwidth,angle=-90,file=1h_rms.ps}}
\centerline{\psfig{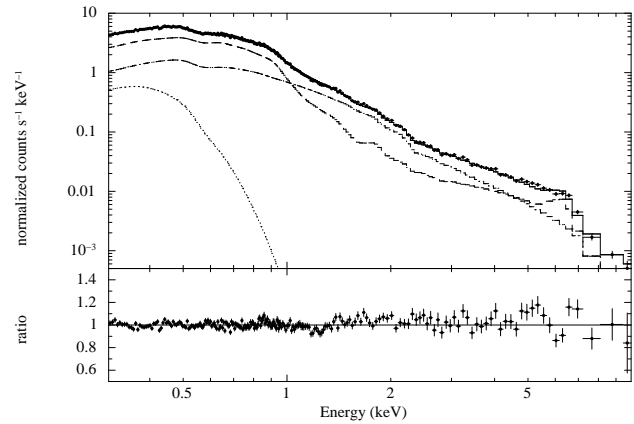}}
\caption{Similar spectrum to Fig.~1 but with an additional reflection
  component (shown summed with the first one) with ionization
  parameters $\xi=528$ and 997. Note that neither this spectrum, nor
  any of the other spectra, has been unfolded. }
\end{figure}

\begin{figure}[h]
\centerline{\psfig{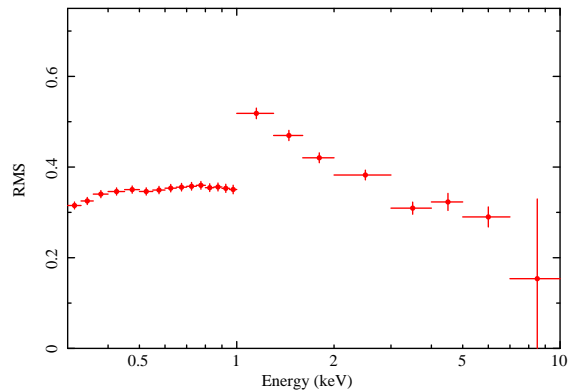}}
%\centerline{\psfig{width=0.5\textwidth,angle=-90,file=reflionx_notcheda.ps}}
\caption{ RMS variability spectrum.  }
\end{figure}

\end{document}